# *Spintronic Functionalities in Multiferroic Oxide-based Heterostructures*

*I. Fina[1] and X. Martí[2]*

[1]*Institut de Ciència de Materials de Barcelona (CSIC), Campus de Bellaterra, 08193 Bellaterra, Spain.*

[2]*Institute of Physics, Academy of Sciences of the Czech Republic, v.v.i., CZ–16253 Praha 6, Czech Republic.*

The list of materials systems displaying both electric and magnetic long range order is short. Oxides, however, concentrate numerous examples of multiferroicity with, in some cases, a large magnetoelectric coupling. As a result, a fruitful research field has emerged contemporaneously with the consolidation of spintronic. The synergy between multiferroics and spintronics was meant to be inevitable and hence the characterization of spintronic functionalities in multiferroic materials is rather abundant. The aim of the present chapter is to review the oxide heterostructures where magnetoelectric coupling is demonstrated by means of spintronic functionalities (i.e. magnetoresistance, anisotropic magnetoresistance, giant magnetoresistance or tunnel magnetoresistance).

# Table of contents

## 1. Introduction

Spintronics [1] is the area that merges electronics with the spin functionalities, and it has been and it is a rapid developing area that has already delivered commercial devices, viz. non-volatile Magnetic Random Access Memory (MRAM) [2]. Spintronics foresees future storage memories, and other information technologies (with enhanced properties) based on spin. Currently, spintronics relies on ferromagnetic materials. In ferromagnets, the magnetic moment can be modified by the application of an external magnetic field, and therefore "0"s and "1"s can be written (**Figure 1**). The two magnetic memory states are antiparallel, and these can be read by means of an adjacent magnetic layer (separated by a nonmagnetic one) with fixed magnetization. The use of this architecture leads to the occurrence of Giant Magnetoeresistance (GMR, Nobel Prize 2007[3]). GMR can be used to electrically-read two magnetic states. Tunnel magnetoresistance (TMR) can be also used to read magnetic information. For TMR, a magnetic tunnel junction architecture, made of two magnetic layers (one fixed and the other containing the magnetic information as in GMR) separated by a thin insulating layer, is used. TMR in magnetic tunnel junctions is consequence of spin dependent tunneling, meaning that depending on their spin state the electrons have very different probability of crossing the thin insulating layer. Thus the difference in the measured resistance defined by the two different magnetic states is larger than in GMR devices. Older and well-known phenomena are magnetoresistance (MR) and anisotropic magnetoresistance (AMR). MR can be ascribed to any variation of the resistance state in a magnetic (or non-magnetic) material under the application of an external magnetic field and its origin can be diverse as it will be discussed in further detail in section 5. AMR is function of the microscopic magnetic moment vector; it is the direction of the spin-axis rather than the direction of the macroscopic magnetization that determines the effect. Therefore, AMR can be used to read perpendicular, instead of antiparallel magnetic states.

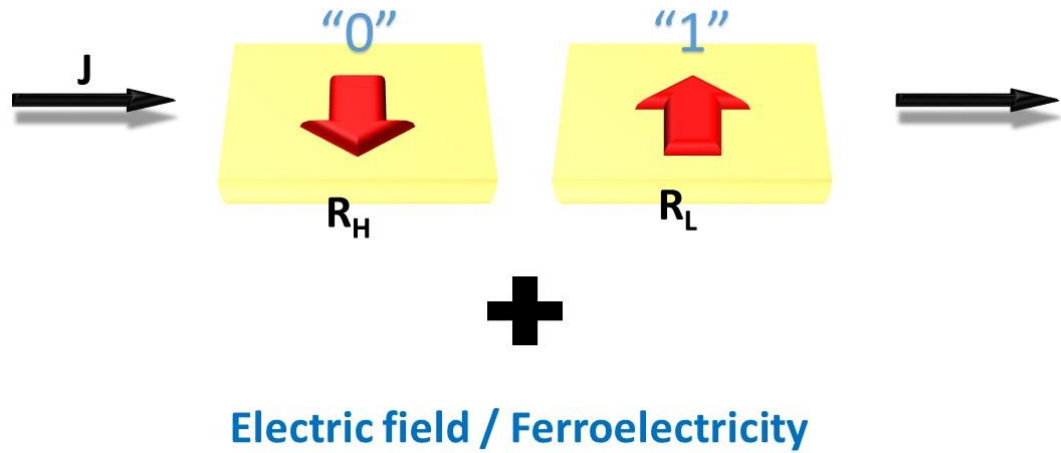

**Figure 1.** In Spintronic based memory devices different magnetic states correspond to different memory states. The read-out of the different magnetization states is easy thanks to the so-called spintronic functionalities (i.e. magnetoresistance, anisotropic magnetoresistance, giant magnetoresistance or tunnel magnetoresistance). Therefore, two different memory states correspond to different resistive states ($R_H$ and $R_L$). The manipulation of magnetization in traditional spintronic devices is achieved by means of the generation of external magnetic field, which is costly in terms of power consumption and dissipation. In some multiferroic materials, where ferromagnetism and ferroelectricity are combined, electric field can be used to modify the magnetic state, avoiding the requirement of large power consumption and dissipation. Multiferroic materials can be read-out by the same spintronic functionalities as in pure ferromagnetic materials.

Recent reports from SIA and SRC, both semiconductor industry associations, point out that the continuous increase of data storage density is reaching its saturation [4]. Hence, the semiconductors industry finds that, an even more important requirement than data density increase, is to improve the energy efficiency. Chappert et al. emphasized in their review entitled *The emergence of spin electronics in data storage* that: "Writing is the problem" [1]. In longitudinal recording systems the magnetization of the recorded bit lies in the plane of the disk. An inductive write element records the data in horizontal magnetization patterns. Alternatives, to simplify and/or decrease the power consumption of the longitudinal recording systems, have been explored. For Hard Disks Device see ref. [5] as a complete review.

The mentioned spintronic effects (TMR, GMR, AMR and MR), which allow the control of electron flow by magnetization, have their reciprocal effect. Thus, the magnetization can be controlled by current injection. This is a recently developed writing technique that does not require the presence of an external applied magnetic field, and it is called the "Spin-Transfer Torque" [6]. Although direct writing by electric current present the convincing advantages of confinement of the switching area and large reduction of consumed energy compared with techniques where an applied magnetic field is required, the amount of consumed energy is still large, and the increase of energy required to keep the temperature of the cell caused by dissipated energy by Joule effect is an issue [7]. For these reasons, there is still a growing interest in finding alternative procedures to process magnetic information and write on spintronic devices using electric fields and fully insulating structures completely avoiding the presence of electric currents. Recent studies have reported that the current can be replaced by electric fields, which can allow to save important amount of energy in magnetic tunnel junctions [8]. However, this method can not avoid the presence of current due to the rather low resistance of MTJ.

Multiferroics are an interesting alternative to pure ferromagnets. Multiferroic materials are those materials where one can find coexistence of more than one ferroic order. Of technological relevance is if one can find coexistence of ferromagnetic (switchable net magnetic moment under application of magnetic field) and ferroelectric (spontaneous surface charge switchable by electric field) order. If coupling between them exists one can envisage the control of surface charge by magnetic field (so-called direct magnetoelectric effect) or the control of magnetization by electric field (so-called converse magnetoelectric effect). As far as the ferroelectric nature of the envisaged material would guarantee its insulating nature, electric currents would be avoided. The absence of currents can lead to an important decrease of the power consumption of the writing procedure. Moreover, the Joule heating issue inherent to the presence of currents will be also avoided resulting in an important decrease of the refrigeration demand of a multiferroic-based electronic element compared with the ferromagnetic-based ones.

Important milestones in the control of magnetic order by electric field have been achieved without the use of oxide compounds [9]. However, the literature is not abundant on the use of materials that are not oxides, signaling the relevant role of oxides in this field.

In oxides, magnetoelectric coupling has been demonstrated [10] in single-phase materials. Cycloidal magnets are an important example of single-phase magnetoelectric materials; however, these are not proper multiferroic materials, because they are antiferromagnetic. Moreover, these only show large effects at low temperatures, making them, in principle, not interesting for applications. The fact that single phase multiferroic/magnetoelectric materials at room temperature are scarce makes composite materials an interesting alternative to them. Composite materials are combination of ferromagnetic and ferroelectric materials at room temperature, therefore the material resulting from their combination must be multiferroic also at room temperature (if structural properties are preserved). Usually, in most of the studied systems, both or one of them is an oxide. In composite materials the coupling is always interface mediated, and several effects can make it possible [11,12].

If multiferroic/magnetoelectric materials are technologically exported, they will make use of a spintronic functionality as a read-out technique. Spintronic probing techniques have been broadly used to probe magnetoelectric coupling. Therefore, it is very relevant to analyze (as we will do in the present chapter) the results obtained up to now on systems where spintronic functionalities are characterized on multiferroic systems. Here we will not only focus on systems that show magnetoelectric coupling, but also on these that, even being multiferroic, does not show coupling. In the present work, we will classify important results published up to now in multiferroics by the participating spintronic functionality:

- **Tunnel magnetoresistance**
- **Giant magnetoresistance**
- **Anisotropic magnetoresistance**

- **Magnetoresistance or resistance manipulation by electric field**

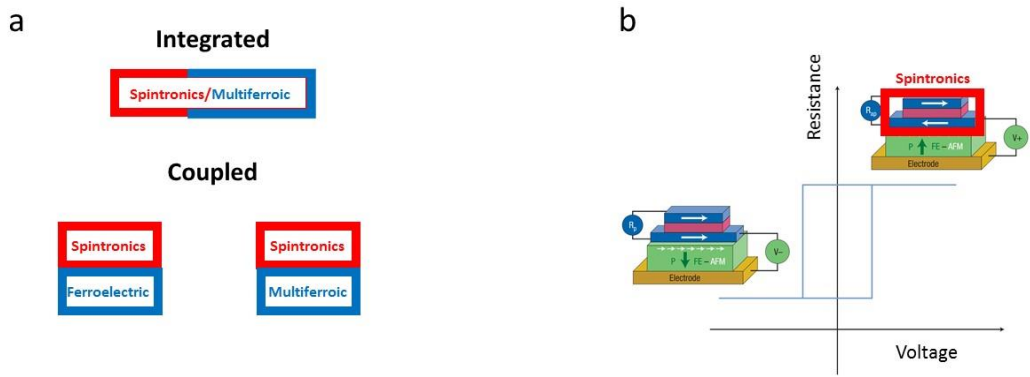

**Figure 2. a.** Classification of multiferroic/spintronic device according to the integration of the multiferroic material in the spintronic device. The upper panel shows the integrated multiferroic-spintronic structure where the multiferroic material is part of the spintronic device. The bottom panels show coupled multiferroic-spintronic structures, where the multiferroic or ferroelectric material is used to manipulate the magnetic state in a spintronic structure. **b.** Sketch of a possible device where the multiferroic material is not an integrated part of the spintronic device. The binary information is stored by the magnetization direction of the bottom ferromagnetic layer (blue) part of a magnetic tunnel junction, which can be electrically read-out through tunnel magnetoresistance. The two parallel (left) or antiparallel (right) states of the magnetic junction can be modified by voltage applied to the multiferroic material ($BiFeO_3$) underneath. The magnetization of the bottom ferromagnetic layer is coupled to the spins in the multiferroic (small white arrows). In the presence of large magnetoelectric coupling in the multiferroic material, the magnetic state of the interface can be modified reversing the ferroelectric polarization, and concomitantly the magnetic and resistive state of the magnetic tunnel junction on top. Adapted from ref. [13].

These four spintronic functionalities can be combined with multiferroicity in different manners. The following classification helps to visualize the level of integration between the spintronic functionality and the multiferroic material:

(i) Integrated: the multiferroic material is at the same time part of the spintronic device, therefore the resistive state would change accordingly with the magnetic state of the multiferroic material [**Figure 2a(top panel)**];

(ii) Coupled: a ferroelectric or multiferroic material changes its electric or magnetic state, and it is somehow coupled with an spintronic device producing also a change on resistance resulting from a change on the magnetic state [**Figure 2a(bottom panel)**].

From this latter group (coupled spintronic-multiferroic structure) is the envisaged oxide-based magnetoelectric random access memories drawn by M. Bibes and A. Barthelemy in ref. [13] in 2008, **Figure 2b**. In ref. [13], $BiFeO_3$ was proposed as active material, and the tunneling device was a passive structure used to read the change on the magnetic state of $BiFeO_3$ induced by electric field. Strictly speaking, $BiFeO_3$ is not a multiferroic material. As the aforementioned cycloidal magnets, $BiFeO_3$ is a robust ferroelectric but it is antiferromagnetic at room temperature. In the envisaged structure, coupling between antiferromagnetic and electric order in the material would be transferred to the tunneling device through magnetic exchange coupling. We will recall $BiFeO_3$ in this chapter due to its relevant role in the field of multiferroics.

In the following sections we will describe the works on multiferroic oxides where tunnel magnetoresistance, giant magnetoresistance, anisotropic magnetoresistance and magnetoresistance are characterized. We will also describe the few works done on spintronic characterization at multiferroic domain walls, which belongs to a particular class of spintronic functionality in a multiferroic systems. We arrange the sections from TMR, the most appealing spintronic functionality, to MR, in principle the less attractive one. Therefore, the reader will see that the degree of complexity throughout sections decreases. In this work, we do not include results obtained by other spintronic functionalities such as Spin Hall Effect, or Anomalous Spin Hall Effect, since the work on these topics reported up to now using multiferroic oxides is scarce.

## 2. *Tunnel magnetoresistance*

In magnetic tunnel junctions metallic ferromagnetic materials acting as emitting and receiving electrodes are spaced by a very thin layer of an insulating non-magnetic material. TMR results from the spin dependent tunneling, whose origin can be found in the splitting of electronic bands for up and down spin states (band scheme of **Figure 3a**). Therefore, the population of carriers for each spin state is different, and the probability to tunnel from the emitting to the receiving electrode in MTJ is different if the magnetic state of the electrodes is antiparallel or parallel (**Figure 3a**). TMR is defined as TMR = $R_{AP}$-$R_P$/$R_P$, where $R_{AP}$ is the resistance state for the magnetic configuration where both electrodes have antiparallel magnetization and $R_P$ is the resistance state for the magnetic configuration where both electrodes have parallel magnetization. The simplest description for the different resistive states is given by Jullier's model that relates the magnitude of TMR with the spin polarization [14].

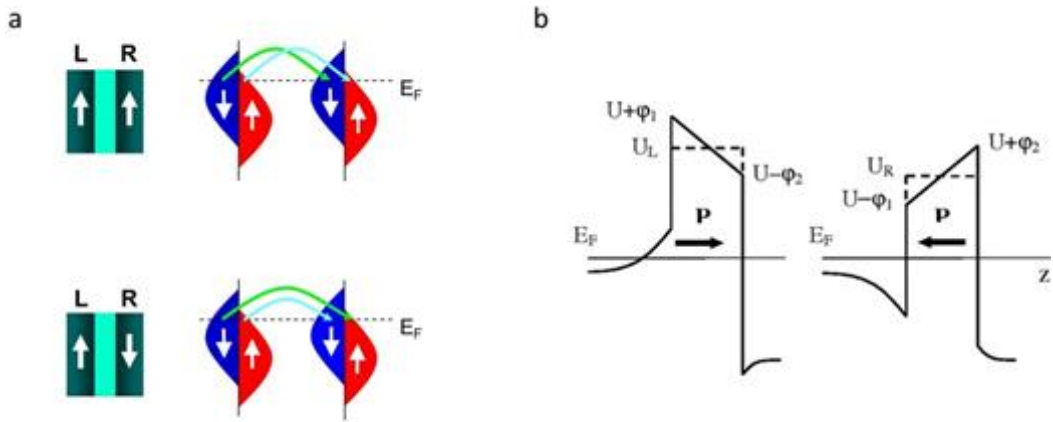

**Figure 3. a.** Schematic representation of the two possible resistive states in a magnetic tunnel junction. In the top panel both electrodes are in parallel configuration and the probability of an electron to tunnel from one to the other is different than the case when their magnetic state is antiparallel, as shown in the bottom panel. Adapted from ref. [15]. **b.** Schematic representation of the potential profile in a Metal-FE-Metal junction for polarization pointing to the left and right, where $\delta/\varepsilon$ ratio is smaller for the metal on the left than for the one on the right. $\varepsilon$ is the metal permittivity and $\delta$ is its screening length. The dashed lines show the average potential for tunneling electrons across the ferroelectric barrier. The horizontal solid line denotes the Fermi energy, $E_F$. Adapted from ref. [16] and corrected as in ref. [17].

Non-equivalent to TMR, but with the similar consequences, is the tunneling electroresistance effect (TER). In a tunnel junction where the insulating spacer is ferroelectric and the two electrodes are not necessarily ferromagnetic but with necessarily different $\delta/\varepsilon$ ratio (where $\varepsilon$ is the metal permittivity and $\delta$ is its screening length), the probability of an electron to tunnel is dictated by the shape of the potential barrier. As sketched in (**Figure 3b**), depending on the direction of the polarization of the ferroelectric materials the barrier profile is different. Therefore, an electrode will have more probability to tunnel if the polarization is pointing to the left than if it is pointing to the right. The scenario and the consequence depicted in (**Figure 3b**) correspond to the situation where the left electrode has smaller $\delta/\varepsilon$ ratio than the right one. TER is defined as TER=$R_\uparrow$-$R_\downarrow$/$R_\downarrow$, where $R_\uparrow$, $R_\downarrow$ correspond to the polarization states for up (right) and down (left) polarization, respectively, although one can also find results in the literature where TER is defined as TER=$R_H$-$R_L$/$R_L$, where $R_H$, $R_L$ correspond to the high and low resistive states, respectively. Even though the concept behind TER effect was proposed time ago [18], it was not until the observation of FE order for very thin ferroelectric films [19-21] that the feasibility of a tunnel ferroelectric barrier was demonstrated.

In the mentioned scenario where the electrostatic effect is governing the change on tunneling current, the correspondence between the resistance state (high or low) and the polarization (up or down) is one to one. Therefore, for the case depicted in **Figure 3b**, the high resistive state corresponds to polarization pointing to the right, and the low resistive state correspond to polarization pointing to the left. However, one can find that literature is not always coherent. This is because the presented scenario is not the only applicable one and other effects might be more predominant or coexisting. The first alternative results from changes on the bonding between atoms upon switching of ferroelectric polarization that can have deep impact on the nature of the electronic configuration at the interface. Thus the effective work-function ($\varphi$) changes, resulting in a change of the electrostatic potential and a concomitant change of the tunneling current. The second alternative is the presence of piezoelectricity. Piezoelectricity [22] is inherent to a ferroelectric materials and it can also result in a change of the actual insulating layer thickness, because of changes on the polar state. However, here if one neglects the presence of electric fields, externally imposed or built-in (as it is the ferroelectric imprint field), the two resistance states would correspond to P=±Ps and P=0, since strain depends on the P absolute value [23,24]. In the presence of an external electric field (easy to find in a FETJ where electrodes with different work-functions are used), the piezoelectric hysteresis loop shifts along the voltage axis resulting in two different strain states at electric remanence. Finally, one must take into account that ionic conduction can also be responsible of important changes on resistance without or partially without any important role of ferroelectric polarization neither, tunneling current [25].

Having stablished the TMR and TER effects, the combination of ferroelectric and ferromagnetic materials or by using a multiferroic material in a tunnel junction architecture results in the so-called multiferroic tunnel junctions, where TER and TMR effects can exist and, in the presence of magnetoelectric coupling, can cross-talk. Multiferroic tunnel junctions can be divided into two different big groups:

- **Single-phase**, where the insulating spacer is multiferroic itself and one or both electrodes are ferromagnetic.
- **Composite**, where the insulating spacer is a ferroelectric or a multiferroic and one or both of the electrodes are ferromagnetic.

In some cases, the spacer is a multiferroic but only its ferroelectric character plays a role on the tunneling effect. In this case we consider that the junction belongs to the second group.

### 1.1. Single phase

The studies on magnetic tunnel junctions where the spacer is a single phase multiferroic material are very limited due to the usually inherent leaky character of multiferroic materials. Leakage current can easily hide the presence of tunneling current. In fact, $BiMnO_3$ is the only material that has shown coexisting presence of TMR and TER [26]. $BiMnO_3$ is multiferroic, but with ferromagnetic and ferroelectric $T_C$ occurring at very low temperature, which limits applications. In **Figure 4a**, it is shown (solid symbols) the TMR curve for a FM/FM-I/non-magnetic tunnel junction (in this case the FM-I is also ferroelectric, LSMO/BiMnO$_3$/Au) after applying appropriate electric prepoling pulses. For prepoling pulses of opposite sign, polarization switches and the resistance changes (as a result of TER) and the overall TMR curve shifts. Moreover, there is a slight difference in the TMR value [27% (after -2 V) and 35% (after +2 V)] indicating some coupling between electric and magnetic states. This effect will be discussed in more detail in the section devoted to TMR on composite multiferroics, where the found effects are larger. Therefore, neglecting the small coupling observed, the tunnel magnetoresistive experiments in BiMnO$_3$ have shown the ability of this material to show 4 resistive states (2-electric and 2-magnetic, **Figure 4b**), being the first demonstration of a multiferroic memory.

Single-phase 4-states multiferroic memory at room temperature might have some interest for some niche of applications. Recent works have shown that there is a short list of available single-phase multiferroic materials at room temperature, in some cases with observed magnetoelectric coupling, ε-Fe$_2$O$_3$ [27,28], Ga$_{2-x}$Fe$_x$O$_3$ [29,30], (Ga,Fe)$_2$O$_3$ [31], Pb(Zr,Ti)O$_3$–Pb(Fe,Ta)O$_3$ [32-34], Pb(Fe, M)$_x$(Zr,Ti)$_{(1-x)}$O$_3$ [M = Ta, Nb] [35], and (1 – $x$)BiTi$_{(1-y)/2}$Fe$_y$Mg$_{(1-y)/2}$O$_3$ – ($x$)CaTiO$_3$ [36]. Nanocomposites, such as BiFeO$_3$-CoFe$_2$O$_4$ [37] or BaTiO$_3$-CoFe$_2$O$_4$ [38], although not being strictly speaking single-phase materials might be also interesting for a 4-states multiferroic memory device. However, reproducing the TMR and TER experiments on BiMnO$_3$ at room temperature remains elusive.

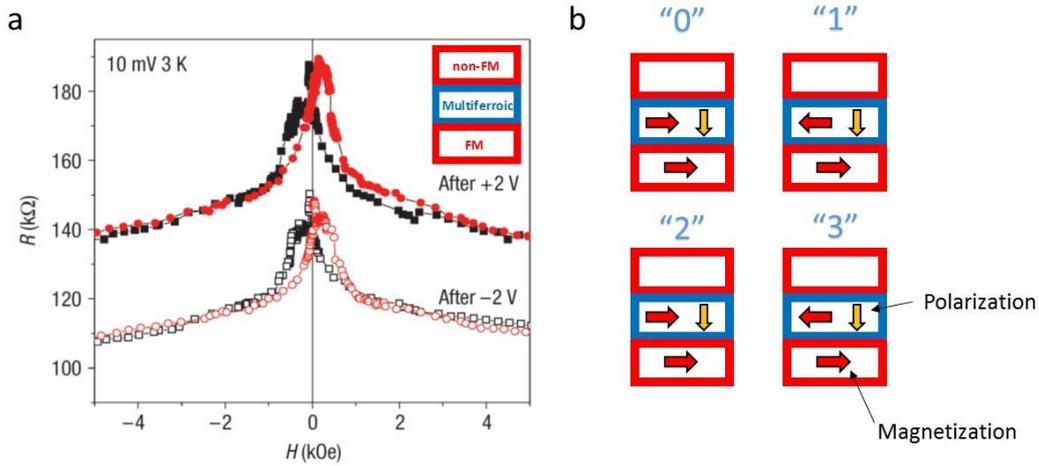

**Figure 4.** Single phase multiferroic tunnel junction. **a.** Tunnel magnetoresistance curves at 4K and 10mV for V = +2V (solid symbols) and −2 V (open symbols) prepoling voltage for LSMO/BiMnO$_3$/Au. Adapted from ref. [26]. **b.** Schematics of the 4 resistive states (2-magnetic and 2-electric) that can be obtained in a multiferroic tunnel junction, where the multiferroic material is the spacer, as in LSMO/BiMnO$_3$/Au junction case.

## 1.2. Composite multiferroics

Now we recall the device sketched in **Figure 2b** where TMR is controlled by electric field. The experimental realization of the mentioned device remains elusive; however, alternative tunnel junctions where multiferroicity is present have been realized and the obtained results are described as follows.

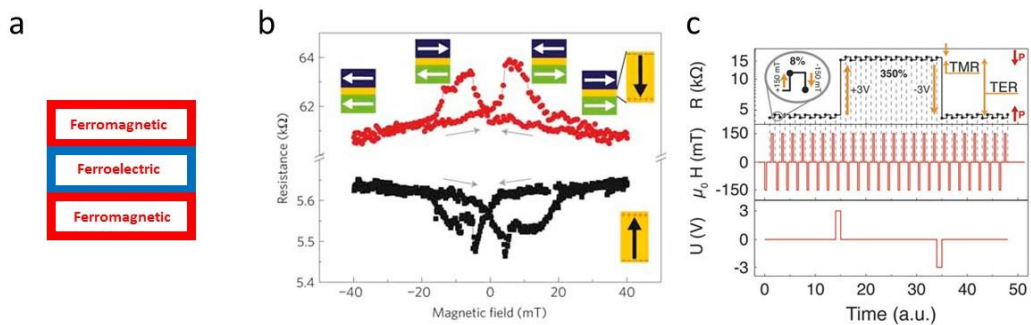

**Figure 5.** Results on (**a**) ferromagnetic/ferroelectric/ferromagnetic tunnel junctions. **b.** TMR for polarization pointing backwards Co (top) and towards Co (bottom) in a LSMO/PZT/Co junction. From ref. [39] **c.** (top panel) Sequentially modified 4 resistive states (magnetic and electric) measured at remanence with magnetic (middle panel) and electric pulses (bottom panel). From ref. [40]. In figure 5, all the experiments are performed at low temperature.

The simplest multiferroic tunnel junction is that where the spacer is ferroelectric and the emitting and the collecting electrodes are ferromagnetic (sketch in **Figure 5a**). In this type of devices, magnetoelectric coupling (if present) takes place at the interface via a list of phenomena: elastic coupling between ferromagenetic/magnetostrictive material and a ferroelectric/piezoelectric one, modification of charge doping by electric field and the concomitant modification in the magnetic ordering, and orbital reconstruction upon ferroelectric switching [11,12]. In fact, it must be stated that in most of the reported systems, all these phenomena must coexist together. Distinguishing between the predominant one might be difficult and in some cases impossible, which superimposes to the different phenomena that can result in TER as mentioned in the introduction. However, the experimental facts that probe magnetoelectric coupling in these systems are irrefutable.

Of interest for us is that the spin polarization (the one that determines the TMR) can be modified thanks to the electrically induced changes of magnetic order. Therefore, TMR value varies with electric field, this is what we have called modulation of TMR (also applicable for GMR, AMR and MR) or tunnel electromagnetoresistance (TEMR), defined as the relative variation of TMR upon polarization switching TEMR = (TMR$_H$ – TMR$_L$)/TMR$_L$, where TMR, TMR correspond to the TMR values for high and low TMR values, respectively. In LSMO/FE/Fe [41,42] or Co [39,40]} structures this fact has been shown. In the LSMO/FE/Fe large negative TMR was found for polarization pointing to Fe, and smaller when pointing away from it. In further investigations [42], it was shown that the change on the spin polarization was finding its origin on the ferromagnetic-like character of BaTiO$_3$, thus providing an interface possessing both magnetic and electric order. Remarkably, when replacing Fe by Co [39], the TMR value (smaller in amplitude) was also negative for polarization pointing to Co, but positive when pointing away from it, thus meaning that the major spin population in one of the electrodes was changing its sign upon ferroelectric switching (**Figure 5b**). A combination of the ferromagnetic-like character antiferromagnetically coupled to one of the ferromagnetic electrodes (Co, in Co/PTO/LSMO structures) was found to be at the origin of large exchange bias effect in multiferroic tunnel junction without requiring any antiferromagnetic or hard magnetic layer, thus producing a multiferroic "spin valve" (**Figure 5c**) [40]. Similar spin-valve effect was reported using an antiferromagnetic layer in LSMO/BTO/Co/IrMn, in this latter case with modulation of TMR [43]. The last example of multiferroic junction where exchange bias is present is the reported LSMO/BFO/LSMO junction, but here the exchange bias effect is low due to the symmetric configuration of the junction [44]. Nonetheless, TMR can also be modified by the electric field in a multiferroic tunnel junction by modifying the electric properties of the insulating spacer. Hambe et al. [45], showed a reversible modulation of TMR from 61 to 69 % depending on the sign of the prepoling voltage. However, in this case, it is argued that ionic displacements are at the origin of the observed modulation of TMR without requiring the presence of magnetoelectric coupling.

Strongly correlated systems, where charge, magnetism and strain are intimately coupled are very attractive from fundamental and applications point of view [46]. The electric field effects on magnetic properties was a rapid emerging field, achieving large effects in a relatively short period of time [47,48]. In recent works, the large electrically induced changes in magnetic/electric properties of manganites have been used to modulate TMR/TER [49,50].

Also by modulating the magnetic properties, and subsequently the transport properties, of one of the ferroelectric/ferromagnetic interfaces in a multiferroic tunnel junction, large pure TER can be obtained. This can be done by modulating the electronic phase in a manganite depending on the ferroelectric polarization. Therefore, the manganite layer can be either metallic or insulating, thus decreasing or increasing the width of the tunneling barrier. This might also result in modulation of TMR [49,50] by electric field, similar to [41], but with different origin. Ferroelectric polarization switch can result also in the modulation of the transport properties in the ferroelectric itself, resulting also in the modulation of the tunneling width. In LSMO/BTO/Pt large values for TER (up to $3 \times 10^4$ %) can be obtained [51]. In Co/PTO/LSMO, the modulation of the tunneling thickness due to metallization of the last layers of the ferroelectric material at the PTO/Co interface was at the origin of large TER (≥230 %) [52].

Now we focus on a particular example of coupled multiferroic magnetic tunnel junctions. It is well-known that in granular magnetic materials the grain boundaries can act as insulating domain walls, resulting in the observation of tunneling current (at very low temperature) in single films of magnetic materials [53-55]. In grainy manganite/FE bilayers, it is a natural argument that the switching of the ferroelectric polarization that results in accumulation or depletion of carriers in the manganite layer can modulate not only its conductivity and/or magnetic state; but also the thickness of the insulating granular boundary thickness. Therefore, an electric modulation of the tunneling barrier height and/or width at the grain boundaries takes place. Thus, electroresistance up to near 1000% can be observed while $T_C$ is being modulated by 16K [56].

As we will show to be also the case for other spintronic functionalities, the characterization of magnetic tunnel junctions grown on top of a piezoelectric material have recently given some interesting results obtaining a larger modulation of TMR by electric field at room temperature [57,58].

Finally, it must be stressed that the device sketched in **Figure 2b** should not be restricted to the use of $BiFeO_3$. Large magnetoelectric coupling between antiferromagnetic and ferroelectric order has been demonstrated in other compounds such as $TbMnO_3$ [10], $TbMn_2O_5$ [59], and other $REMnO_3$ (RE=rare earth) compounds [60-63], among other more complex and more recently studied oxides [64]. In these materials the coupling is intrinsic, thus the ferroelectric polarization appears thanks to a particular magnetic order, a cycloidal one. However, we find two reasons that prohibit their application. First, the cycloidal order results from magnetic frustration and therefore it can only appear at very low temperature. Second, the fact that there is not any reported clear procedure to exploit this magnetoelectric coupling in a structure where cycloidal order is somehow coupled to a ferromagnetic order in an adjacent ferromagnetic layer.

## 3. *Giant magnetoresistance*

If in a magnetic tunnel junction, one replaces the insulating spacer by a non-magnetic conductive material, the new device architecture is that of a GMR. In the GMR architecture, it is important to distinguish between two subgeometries: the transversal and the longitudinal ones. In the transversal (or current perpendicular to the plane, CPP) the current is perpendicular to the spacer plane. Therefore, if the magnetic alignment of the emitting and receiving electrode is parallel the resistance is low, and if it is antiparallel the resistance is high (in the archetypical case of positive GMR). As most of the current is scattered at the interface between layers, in the longitudinal geometry (or current in plane, CIP) GMR is also present; however, as far as the current is parallel to the interface, the efficiency is lower. The former has the disadvantage of being more difficult to grow, being that the reason why in all the multiferroic structures showing GMR the latter architecture has been used. As in TMR, GMR is defined as GMR = $R_{AP}-R_P/R_P$, where $R_{AP}$ is the resistance state for the magnetic configuration where both metals have antiparallel magnetization and $R_P$ is the resistance state for the magnetic configuration where both metals have parallel magnetization.

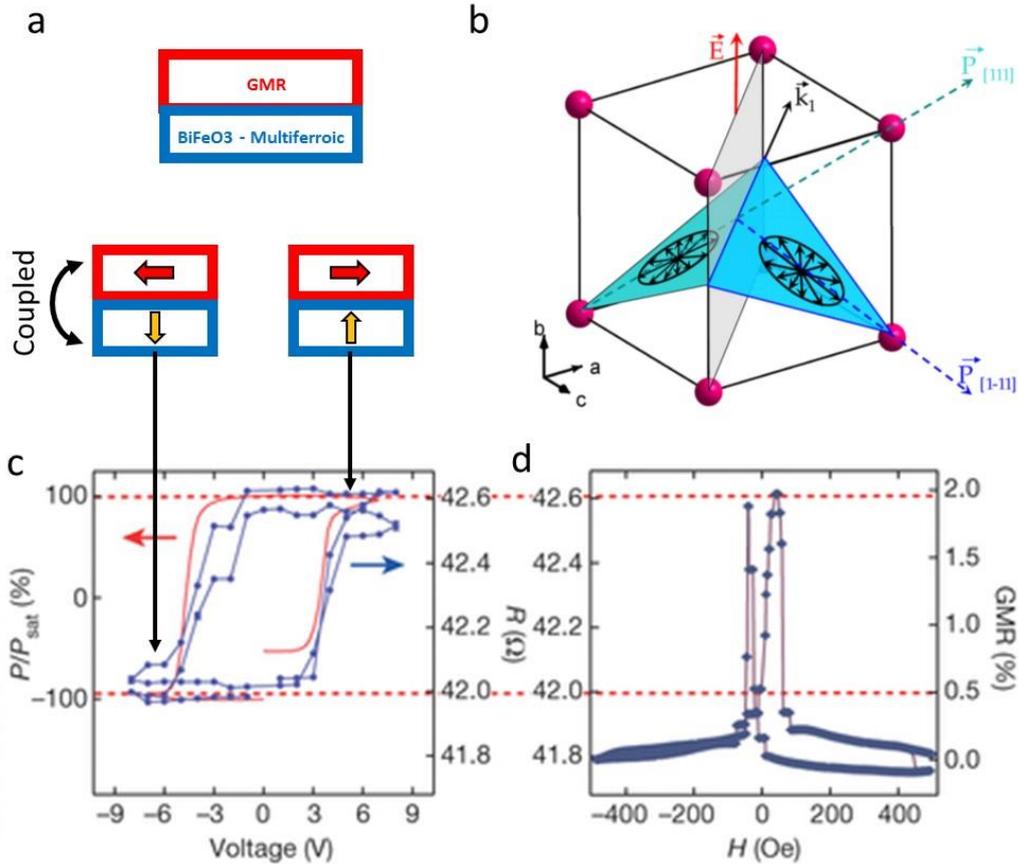

**Figure 6. a.** The magnetic state of the free layer of the GMR structure (red) is coupled to the polarization of BiFeO$_3$, therefore if polarization on BiFeO$_3$ is manipulated the magnetic state changes and the resistive state in the GMR structure is modified accordingly. **b.** Schematics of the rotation of the cycloidal plane accompanying the change of the polarization direction. From ref. [65]. **c.** Change of polarization (red line) and resistance (blue points) as a function of the applied electric field in a BFO/CoFe/Cu/Co structure. **d.** GMR in the CoFe/Cu/Co structure grown on top of BFO. From ref. [66].

In the GMR architecture there is no insulating layer involved. This makes impossible to have a GMR where the multiferroic materials is part of it, since ferroelectricity only exists in insulating materials. Therefore, GMR architecture can not be an integrated multiferroic-spintronic device. However, the rapid developing studies on metallic or semiconducting ferroelectrics might allow it, at least from fundamental point of view (see f.i. [67]). Therefore, the study of GMR in multiferroic devices is limited to the electric manipulation of GMR in a coupled spintronic-multiferroic structure, similar to that one introduced in the previous section and introduced in ref. [13], replacing the magnetic tunnel junction by a GMR structure.

In BiFeO$_3$/FM multiferroic structures the most relevant results have been obtained. The fact that BiFeO$_3$ is not ferromagnetic makes the participation of a coupled ferromagnetic material and magnetoelectric coupling necessary factors for the envisaged device. In the presence of these necessary factors GMR is an excellent option to probe changes on magnetic order (**Figure 6a**). In early works, it was already demonstrated the presence of coupling between magnetism and applied electric field [68,69]. Afterwards, it was shown that the cycloidal order superimposed to the collinear antiferromagnetic order and the ferroelectric order were coupled [65] (**Figure 6b**). Owing to the presence of this type of coupling, it was demonstrated that the magnetic order in an adjacent to BiFeO$_3$ ferromagnetic layer was can be rotated by application of electric field in BiFeO$_3$ [66] (**Figure 6c**). In previous experiments AMR [70] was used to probe magnetic order, and therefore the experiments will be discussed in more details in the pertinent section. Remarkably, it was reported that the observed electrically stimulated GMR contrast, although small ($\approx$1.5%), was almost the same that the one obtained while sweeping the magnetic field (compare **Figure 6c,d**) and thus reversing the magnetization by 180º. The small pure GMR was expected due to the used GMR device was a longitudinal one, which is known to be much less effective than the vertical one. However, the fact that both, electrically and magnetically induced changes on resistance were similar, indicates that coupling was very efficient. This was argued to be owing to the particular ferroelectric switching procedure used.

Other systems where the magnetoelectric coupling is purely strain mediated or is argued to be strain mediated have been studied, since the early publication on PZT/Spin-Valve-GMR structure [71]. The fact that all the characterization in piezoelectric/GMR systems was performed in CIP configuration, limited the obtained variation of GMR induced by electric field to small numbers (<1%) [72-74].

## 4. *Anisotropic magnetoresistance*

Anisotropic magnetoresistance (AMR) was discovered by Lord Kelvin more than 100 years ago in a piece of iron [75]. AMR results from the change of the conductivity a ferromagnetic material depending on the direction of magnetization with respect the injected current used to measure the resistance. It results that the resistivity (ρ) depends on the angle (θ) between the current and the applied magnetic field as: $\rho(\theta) = \rho_\perp + (\rho_\parallel - \rho_\perp) \cos 2\theta$ [76]; thus, AMR is an odd function of the applied magnetic field and can thus not distinguish between antiparallel states, but only between perpendicular ones. The AMR amplitude is small in front of the amplitude that one can get using TMR or GMR architectures; however, it presents the enormous advantage of being a very simple technique. Similar to GMR, AMR can not exist in multiferroic materials because of its insulating nature and therefore we will focus on coupled spintronic-multiferroic structures, as shown in the top the sketches of **Figure 7**.

Seminal work on the modulation of AMR by electric field was that shown by the characterization of Pt/YMnO$_3$/Py sandwich structures [77]. In this work the electric control of antiferromagnetic order in YMnO$_3$ (due to domain wall coupling between ferroelectric and antiferromagnetic domain walls as predicted by Goltsev et al. [78]) and consequently the exchange bias effect between Py and YMnO$_3$ allowed to strongly tune the AMR of the Py layer (**Figure 7a**).

In manganite/ferroelectric structures, modulation of the longitudinal and transversal AMR, depicted in the figures is shown ref. [47,79] (**Figure 7b**). In this latter case the field induce change on doping state of the manganite films, resulting in an anisotropy change that modifies the observable AMR.

Important changes of AMR by electric field are those obtained by J.T. Heron et al. in BFO/Co structures [66]. On BFO/Co, as shown by **Figure 7c**, the AMR was shifting by 180º at room temperature. In the experiment, AMR was used to probe the alignment of the magnetic moment. In the experiment AMR measurements are performed at very low magnetic field, the observed cos2θ can become cosθ due to pinning of magnetic moment that only allow the magnetization to slightly move around its pinned position. Therefore, opposite magnetic states can be distinguished by a shift of 180° in the AMR data, as shown in **Figure 8c**, meaning that the modulation was -100 %. It was argued the 180° magnetic reversal took place thanks to the coupling between the canted BFO moment and the ferroelectric polarization, which at the same time was coupled to Co.

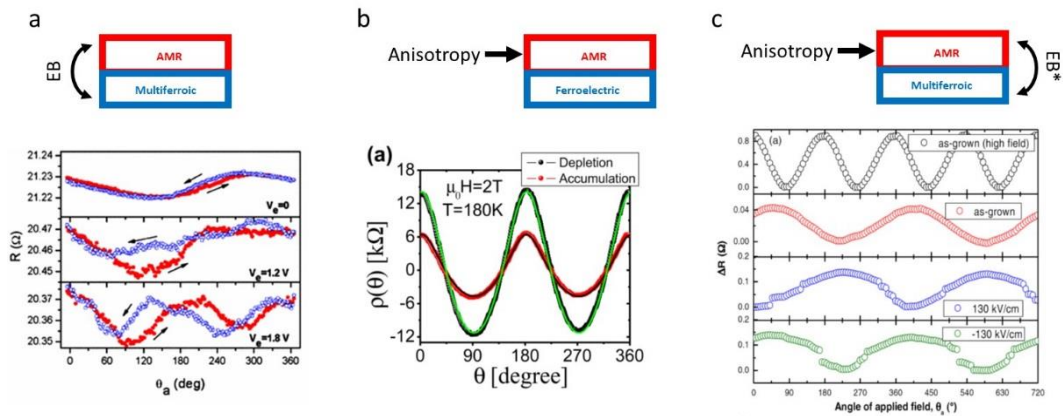

**Figure 7. a**. Electric control of AMR in YMnO$_3$/Py structures, in which coupling is mediated by magnetic exchange coupling. From ref. [77]. **b**. Electric control of transversal AMR in La$_{0.825}$Sr$_{0.125}$MnO$_3$/PZT heterostructures where, upon ferroelectric switching, magnetic anisotropy of the manganite changes. From ref. [79]. **c.** Electric control of AMR in BFO/Co structures. From ref. [70]

As in the case of tunnel magnetoresistance, results on the electric modulation of AMR has been reported by the use of a piezoelectric material of large piezoelectric coefficient and magnetoelectric coupling mediated by strain. In ref. [80], Hong et al. showed a sizeable but small modulation of AMR in a Ni/SiO$_2$/Ti/(011)-PMN-PT heterostructure.

## 5. *Magnetoresistance or electric field modulation of resistance and magnetism*

Magnetoresistance is defined as the change of the resistivity of a material by the application of an external magnetic field. Magnetoresistance itself can have several different origins. The first one is that resulting from the action of Lorentz force and it takes place in any magnetic or non-magnetic metallic material. In ferromagnetic materials its origin can be also due to grainy magnetoresistance, or anisotropic magnetoresistance. In grainy magnetoresistance the conductivity of the material changes according to the number of interfaces with different magnetic moment (domain walls), where stronger scattering exists. Therefore, one can find the maximum variation of resistance at the coercive field, where the number of domains and domain walls is maximum. In AMR magnetoresistance the change on resistance results from the misalignment of some or all magnetic domains with the reductions of applied magnetic field, which is favorable in very isotropic medias. Therefore, the angle between the injected current and the magnetic moment is in average non-zero and the conductivity increases or decreases accordingly. Finally, magnetoresistance can also appear owing to the presence of spin-orbit coupling, the conductivity in this case is modified by any change on the magnetic moment or magnetic order.

Therefore, transport measurements are directly related to the magnetization. This has allowed also to infer changes on transition temperature in multiferroic structures. In ferroelectric/ferromagnetic semiconductor devices, where the coupling is mediated by electric field (ferroelecric field effect devices), $T_C$ has been tuned by 5-10 K degree. Fully oxide structures, with a strongly correlated material (SCM) [81] acting as ferromagnetic showed the largest modulation on the magnetic properties induced by ferroelectric polarization reversal [48,82]. The observed shift of the Curie temperature ($T_C$) for the ferroelectrically-gated SCM is much larger (by 40 K) than the one measured for a FM-SC channel (compare, for instance, refs [83] and [82]), and the effects are still large (by 16 K) in polycrystalline SCM/ferroelectric structures grown on Si. Moreover, $T_C$ values characterizing the SCMs are larger than in FM-SCs [56].

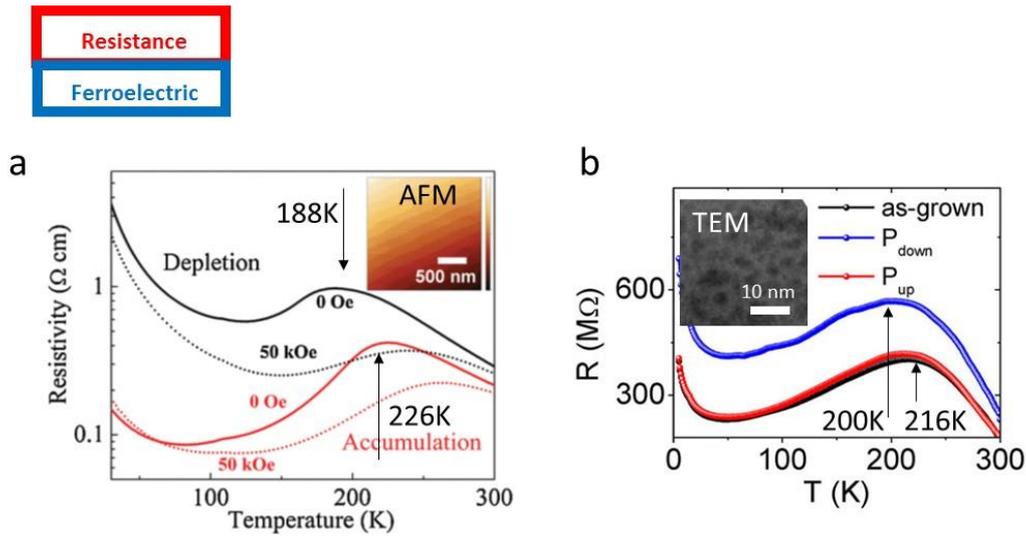

**Figure 8. a.** Resistance versus temperature measured upon polarization switching inducing the accumulation/depletion of carrier at the interface in a $PbZr_{0.2}Ti_{0.8}O_3/La_{0.8}Sr_{0.2}MnO_3/SrTiO_3(001)$ fully epitaxial heterostructure structure. Good quality of the film is shown by the AFM image included as an inset. From ref. [48]. **b.** Equivalent measurement than in **a**, in a fully polycrystalline $PbZr_{0.2}Ti_{0.8}O_3/La_{0.825}Sr_{0.175}MnO_3$ structure grown on Si. In the inset it can be observed the polycrystalline character of the sample revealed by a TEM image. From ref. [56].

In FeRh/piezoelectric structures the measurement of the resistance was used to probe that owing to strain mediated coupling complete modulation of the antiferromagnetic-to-ferromagnetic transition was achieved by electric field when using $BaTiO_3$ as a piezoelectric material [84], instead of the partial one obtained in the case that the used piezoelectric material was PMNPT [85].

In pioneering multiferroic structures, standard measurement of resistance has been used to measure important variations of magnetization [86] where the magnetoelectric coupling was mediated by a field effect. The fact is that the measurement of the resistive state versus electric field can be used as a tool to probe magnetism and give interesting insight on the physical mechanism triggering magnetoelectric coupling [87]. In **Figure 9a**, it is shown that the resistance in a $La_{0.7}Sr_{0.3}MnO_3$/PZT structure is a hysteretic function of the applied electric field at remanence, thus reproducing the typical ferroelectric loop like shape of the ferroelectric underneath. Therefore, in this case the coupling is mediated by field effect induced by the ferroelectric that modifies the charge density in the ferromagnet, and concomitantly its transport and magnetic properties. Instead in **Figure 9b**, similar structure $La_{0.8}Ca_{0.2}MnO_3$/PZT shows butterfly shape piezoelectric loop. Thus, in this latter the coupling is elastic owing to the piezoelectric nature of the ferroelectric and the magnetostrictive effect of the ferromagnet.

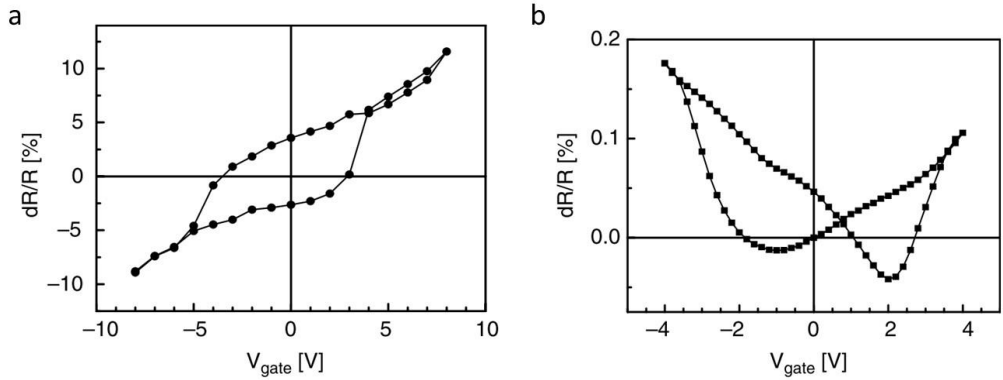

**Figure 9.** a. Resistance modulation dR/R vs. piezoelectric voltage in a $La_{0.8}Ca_{0.2}MnO_3$ (10 nm)/PZT structure, with dominating field effect. b. The same experiment than in a. in $La_{0.7}Sr_{0.3}MnO_3$/PZT, where strain effect dominates. From ref. [87].

Direct experiments focused on the electric control of magnetoresistance are scarce [88]; because its measurement add little information on the simple measurement of resistance or temperature dependence on resistance, and their magnitude is low.

## 6. *Spintronic functionalities at BiFeO₃ domain walls*

Enhanced conductivity at BiFeO3 domain walls has been reported using proximity probe techniques [89]. However, the difficulties found on the realization of microcontacts to explore its temperature and other parameters dependency limited their characterization to room temperature ambient conditions characterization. Li doping in BFO allowed large improvement of insulating properties in $BiFeO_3$. Therefore, allowing the characterization of transport properties dependence on temperature. In reference [90], AMR at the domain walls revealed that sizeable response can be observed and indicating their ferromagnetic character. Interestingly, the result showed a large hysteresis depending on the increase or decrease of the angle of the current with respect to the magnetic field. This was argued to result from the magnetic exchange produced between the ferromagnetic domain wall and the antiferromagnetic domains of $BiFeO_3$. AMR was already observed in ref. [91] at room temperature, but in this latter case using as-grown in-plane domain walls present at $BiFeO_3$ and in-plane contact geometry.

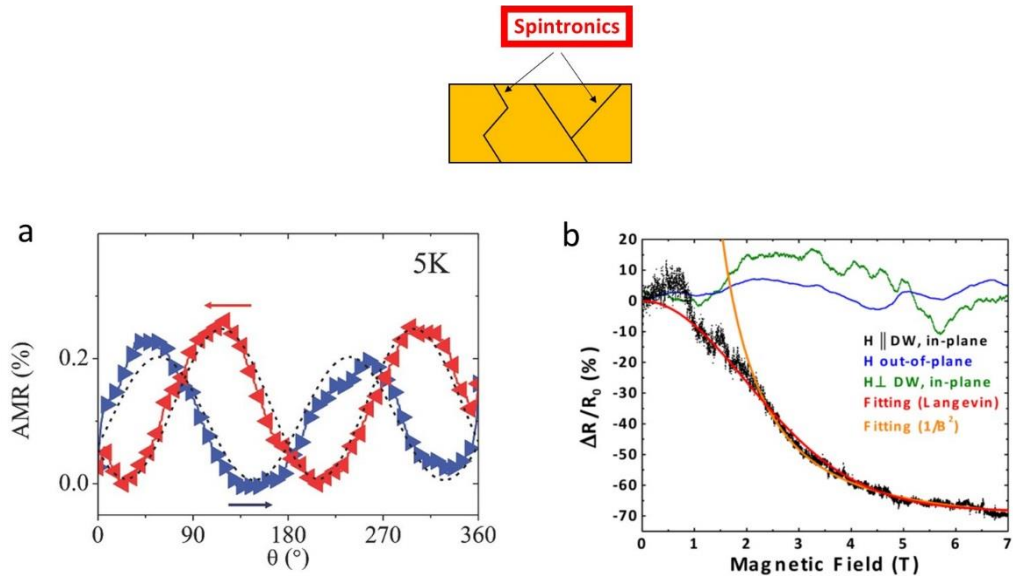

**Figure 10. a.** AMR measured at $BiFeO_3$ domain walls. From ref. [90]. **b.** MR for applied magnetic fields along different direction with respect the as-grown BiFeO3 domain walls, showing the presence of AMR. From ref. [91].

## 7. Conclusions and perspectives

The tremendous effort done by the scientific community on the characterization of multiferroic materials has led to several results, important from the fundamental and technological points of view. In

**Table** 1, the results obtained in multiferroic systems where any spintronic functionality (TMR, AMR, GMR, MR) has been reported are summarized. Regarding TMR, we can conclude that there is a lack of work performed in systems that can work at room temperature. This is mainly because the multiferroic materials/structures so far employed lose their multiferroic character at room temperature, and electric or magnetic state dependent tunneling only appear at low temperature. The exceptions are structures where the magnetic tunnel junction is not a multiferroic itself, in particular the MTJ/piezoelectric structures, where electric control of TMR has been observed. The lack of results at room temperature reported for TMR is not present fot GMR and AMR, where interesting results have been obtained in BFO-based structures. The absolute variations of resistivity in these latter systems is always low due to the low GMR values in used planar configuration and also the low values of AMR. Also modulation of GMR and AMR at room temperature has been achieved in GMR/piezoelectric and AMR/piezoelectric structures similar to TMR/piezoelectric systems. In the spintronic/piezoelectric systems studied up to now the piezoelectric material is a single-crystal with obvious disadvantage for applications. Regarding MR, because of the aforementioned reasons, the works are also limited, which makes difficult to extract any conclusion.

Therefore, two are the open venues regarding the electric control of spintronic functionalities:

- To expand the knowledge on magnetoelectric coupling in BFO-based or similar structure to achieve electric control of magnetic order that produces larger resistance variations in a coupled spintronic-multiferroic.
- To study the viability and the applicability of spintronic/piezoelectric systems, in particular how the observed effects can be also preserved after nanostructuration, as required for miniaturized devices.

Finally, one should remark that the fact that AMR and MR were observed in BFO domain walls opens the door to new unknown applications, where the electric manipulation of spintronic nano-objects is necessary.

| Material | TER | ER | Modulation | TMR | GMR | AMR | MR | Temperature | Ref |
|---|---|---|---|---|---|---|---|---|---|
| LSMO/BMO/Au | 27% | | 30% | 30% | | | | 3K | [26] |
| LSMO/BFO/LSMO | 45% | | 13% | 69% | | | | 80,0 | [45] |
| LSMO/BTO/Fe | 16% | | 467% | 45% | | | | 4,2K | [41] |
| LSMO/PZT/Co | 1000% | | -3693% | 10% | | | | < 200K | [39] |
| LSMO/BTO/LCMO | 8000% | | ∞ | 85% | | | | < 100K | [49] |
| LSMO/PTO/Co | 350% | | 33% | 8% | | | | < 140K | [40] |
| LSMO/BTO/Co/IrMn | 10000% | | 100% | 20% | | | | 10K | [43] |
| LSMO/BFO/LSMO | 15% | | 18% | 3% | | | | 10K | [44] |
| PMNPT/CoFeB/AlO/CoFeB | | 15% | # | 40% | | | | RT | [57] |
| PMNPT/Ta/CofeB/MgO/CoFeB/Ta | | 100% | # | 100% | | | | RT | [58] |
| PZT/Ta/IrMn/FeCo/Cu/NiFe/Ta | | $ | 15% | | 2% | | | RT | [71] |
| BFO/CoFeB/Co | | $ | 40% | | 80% | | | RT | [72]* |
| PZT/Pt/IrMn/Cu/Co/Cu/CoFeB/MgO | | 0,02% | ≈0% | | 0,4% | | | RT | [73] |
| BFO/CoFe/Cu/CoFe | | 1,5% | $ | | 2% | | | RT | [66] |
| BTO/Fe/Cu/Co | | $ | 25% | | 2,5 | | | RT | [74] |
| YMO/Py | | 100% | 275% | | | 0,1% | | 5K | [77] |
| BFO/CoFeB | | -100% | -100% | | | 0,1% | | RT | [70] |
| LSMO/PZT | | 50% | 50% | | | 1% | | < 250K | [79] |
| PMNPT/Ti/SiO2/Ni | | $ | 0,15% | | | 1% | | RT | [80] |
| PZT/LSMO(poly) | | 1000% | 33,0% | | | | 50% | <200 K | [56]* |
| PMNPT/Co | | 0,15 | ≈0% | | | | 0,4% | RT | [88] |
| BFO domain walls | | 400% | ≈0% | | | 0,2% | | 5K | [90] |
| BFO domain walls | | $ | $ | | | $ | | 10K | [91] |

TMR, AMR and MR highest value if there is modulation

#Important shape modulation without change on GMR amplitude

*irreversible

$ Data not available

RT: room temperature

Table 1. Survey of the TER, ER, TMR, GMR, AMR, MR results obtained in several multiferroic systems based on oxide materials. The modulation is the variation of the Spintronic functionality (SF=TMR,GMR,AMR,MR) upon the application of an electric field. Modulation=(SF$_{high}$-SF$_{low}$)/SF$_{low}$).

## 9.     Acknowledgments

We acknowledge financial support from the Spanish MINECO (MAT2015-73839-JIN). We also acknowledge support from the ERC Advanced grant no. 268066, from the Ministry of Education of the Czech Republic Grant No. LM2015087, from the Grant Agency of the Czech Republic Grant no. 14-37427. ICMAB-CSIC authors acknowledge financial support from the Spanish Ministry of Economy and Competitiveness, through the "Severo Ochoa" Programme for Centres of Excellence in R&D (SEV- 2015-0496). I.F. acknowledges Beatriu de Pinós postdoctoral scholarship (2011 BP-A_2 00014) from AGAUR-Generalitat de Catalunya and Juan de la Cierva – Incorporación postdoctoral fellowship (IJCI-2014-19102) from the Spanish Ministry of Economy and Competitiveness of Spanish Government. J. Sort and M. Qian are acknowledged for critical reading.